\documentclass[10pt,pra,twocolumn,floatfix]{revtex4}
\usepackage{amssymb}
\usepackage{amsfonts}
\usepackage{amssymb}
\usepackage{amsmath}
\usepackage[dvips]{graphicx}

\begin{document}

\title{Swapping of Gaussian Einstein-Podolsky-Rosen steering}
\author{Meihong Wang$^{1}$, Zhongzhong Qin$^{1,2}$ and Xiaolong Su$^{1,2,}$}
\email{suxl@sxu.edu.cn}
\affiliation{$^{1}$State Key Laboratory of Quantum Optics and Quantum Optics Devices,
Institute of Opto-Electronics, Shanxi University, Taiyuan, 030006, People's
Republic of China \\
$^{2}$Collaborative Innovation Center of Extreme Optics, Shanxi University,
Taiyuan 030006, China \\
}

\begin{abstract}
Einstein-Podolsky-Rosen (EPR) steering is a quantum mechanical phenomenon
that allows one party to steer the state of a distant party by exploiting
their shared entanglement. It has potential applications in secure quantum
communication. In this paper, we present two swapping schemes of Gaussian
EPR steering, single-channel and dual-channel schemes, by the technique of
entanglement swapping. Two space-separated independent EPR steering states
without a direct interaction present EPR steering after deterministic
swapping. By comparing the EPR steering of the single-channel and
dual-channel schemes, we show that the transmission distance of the
single-channel scheme is much longer than that of the symmetric dual-channel
scheme. Different from entanglement swapping, one-way EPR steering is
presented after swapping over lossy channels. The most interesting thing is
that the change of the EPR steering direction is observed in the
dual-channel scheme. We also show that excess noise in a quantum channel
will shorten the transmission distance of the swapping, even leading to the
sudden death of EPR steering. The presented schemes provide a technical
reference for remote quantum communications with EPR steering.
\end{abstract}

\maketitle

\section{Introduction}

Einstein-Podolsky-Rosen (EPR) steering was first noted by Einstein,
Podolsky, and Rosen in their famous 1935 paper \cite{EPR}. They argued the
completeness of quantum mechanics by calling this phenomenon
\textquotedblleft spooky action at a distance." In his response to the EPR
paper, Schr\"{o}dinger originally introduced the concept of EPR steering 
\cite{Schrodinger1,Schrodinger2}. Suppose Alice and Bob share an EPR
entangled state and they are separated in space. EPR steering means that one
party, say, Alice, can \textquotedblleft steer" the state in Bob's station
by performing a measurement on her state at a distance, i.e., if Alice makes
a measurement on her state, the state in Bob's station will change
instantaneously. In the hierarchy of quantum correlations, EPR steering
represents a weaker form of quantum nonlocality and stands between Bell
nonlocality \cite{Bell} and EPR entanglement \cite{EPREntanglement}.
Concretely, the violation of Bell inequality implies EPR steering in both
directions, and EPR steering of any direction implies that the quantum state
is entangled \cite{WisemanPRL}.

EPR steering has recently attracted increasing interest in the quantum
optics and quantum information communities \cite%
{WisemanPRL,WisemanPRA,Quantifying}. EPR steering can be regarded as
verifiable entanglement distribution by an untrusted party, while entangled
states need both parties to trust each other, and Bell nonlocality is valid
assuming that they distrust each other \cite{WisemanPRA}. In the field of
quantum information processing, EPR steering has potential applications in
one-sided device-independent quantum key distribution \cite{QKD}, channel
discrimination \cite{ChannelDiscrimination}, and teleamplification \cite%
{teleamplification}.

The inherent asymmetric feature is the unique property of EPR steering that
differs from both entanglement and Bell nonlocality. There are situations
when Alice can steer Bob's state but Bob can not steer Alice's state, or
vice versa, which are referred to as one-way EPR steering \cite{WisemanPRL}.
The demonstration of one-way EPR steering is of foundational significance in
testing the basic laws in quantum mechanics and has potential applications
in asymmetric quantum information processing. Gaussian one-way EPR steering
has been demonstrated with a two-mode squeezed state \cite{OneWayNatPhot}
and a multipartite EPR steering system \cite{OneWayPKLam}, with their
measurements restricted to Gaussian measurements. Other measurement methods
used to show the property of one-way EPR steering have been theoretically
constructed, including general projective measurements \cite{OneWayTheory1},
arbitrary finite-setting positive-operator-valued measures (POVMs) \cite%
{Quantifying}, infinite-setting POVMs \cite{OneWayTheory2}, and infinite
number of arbitrary projective measurements \cite{OneWayTheory3}. Very
recently, genuine one-way EPR steering was experimentally demonstrated by
two groups independently \cite{OneWayPryde,OneWayGuo}, based on proposals in
Refs. \cite{OneWayTheory3} and \cite{OneWayTheory1}, respectively.

In a quantum network, the remote transfer of a quantum state is an important
step in quantum communication. Entanglement swapping \cite%
{Zukowski1993,Ralph,Tan1999,Loock1999,Zhang2002}, which makes two
independent quantum entangled states without a direct interaction become
entangled, is an important technique in building a quantum information
network \cite{Duan2001}. In fact, it represents the quantum teleportation of
an entangled state \cite{Loock1999}. Entanglement swapping has been
experimentally demonstrated in both discrete and continuous-variable regions 
\cite{Pan1998,Jia2004}. Recently, entanglement swapping between discrete and
continuous-variable systems has been demonstrated \cite{Takeda}.
Entanglement swapping among three two-photon EPR entangled states has been
used to generate a Greenberger-Horne-Zeilinger state \cite{Lu2009}. The
technique of entanglement swapping has been applied to complete the remote
transfer of Gaussian quantum discord \cite{Lingyu Ma2014}. Very recently,
quantum entanglement swapping between two multipartite entangled states has
been demonstrated experimentally \cite{Su2016}, which shows the feasibility
of connecting two multipartite entangled states by entanglement swapping.

In this paper, we apply the technique of entanglement swapping to realize
the deterministic swapping of Gaussian EPR steering. We show that the
steering property exists between two independent states without a direct
interaction after entanglement swapping. Specifically, two kinds of swapping
schemes are compared, which are called the single-channel scheme and
dual-channel scheme, respectively. We theoretically analyze the swapping of
EPR steering in a realistic environment, where the optical modes are
transmitted in one or two lossy and noisy channels. The dependence of EPR
steering on transmission distances and excess noise in quantum channels is
presented. The results show that the transmission distance of a
single-channel scheme is much longer than that of a symmetric dual-channel
scheme. The most interesting result is that the change of the EPR steering
direction is observed in the dual-channel scheme, which is a phenomenon
related to the asymmetric property of the EPR state. The effect of excess
noise in a quantum channel on output EPR steering is also analyzed. We show
that excess noise can shorten the transmission distance of the swapping
scheme and lead to the sudden death of EPR steering. The presented results
provide a realistic reference to construct a quantum communication network
with EPR steering.

\section{EPR steering swapping schemes}

Figure 1 shows the schematic of EPR steering swapping. Alice and Bob own two
independent EPR entangled states $\left( \hat{A},\hat{B}\right) $ and $%
\left( \hat{C},\hat{D}\right) $, which have the property of EPR steering,
respectively. There is no direct interaction between these two
space-separated EPR states. In order to establish EPR steering between them,
we apply the swapping scheme to them. We consider two kinds of swapping
schemes, i.e., a single-channel scheme and dual-channel scheme, where the
state is transmitted over a single quantum channel and two quantum channels,
respectively.

In the single-channel scheme, as shown in Fig. 1(a), Alice sends mode $\hat{B%
}$ of her EPR state to Bob through a quantum channel. Bob performs a joint
measurement on the received optical mode $\hat{B}$ and one of the EPR mode $%
\hat{C}$ hold by himself. The joint measurement is performed by coupling
them on a 1:1 beam splitter and measuring the amplitude and phase
quadratures of the output modes $\hat{E}$ and$\ \hat{F}$ by two homodyne
detectors (HDs), respectively. The measurement results are fed forward to
mode $\hat{D}$ by classical channels. Bob performs phase-space displacement
on quantum mode $\hat{D}$ with amplitude and phase modulators, respectively.
In the dual-channel scheme, as shown in Fig. 1(b), Alice and Bob send one of
each state ($\hat{B}$ and $\hat{C}$) to the middle station owned by Claire
through two quantum channels, respectively. Claire performs a joint
measurement on the received optical modes $\hat{B}$ and $\hat{C}$. The
measurement results are fed forward to mode $\hat{D}$ through classical
channels. Finally, EPR steering between $\hat{A}$ and $\hat{D}^{\prime }$ is
verified.

In the dual-channel scheme, when the transmission distances of quantum
channels $T_{1}$ and $T_{2}$ are different, it corresponds to an asymmetric
swapping scheme. For the asymmetric swapping scheme, the middle station is
placed near Alice's or Bob's station. The single-channel scheme is a special
case of the asymmetric swapping scheme. When the transmission distances of
quantum channels $T_{1}$ and $T_{2}$ are the same, it corresponds to a
symmetric dual-channel scheme. In this case, the distances from the middle
station to Alice's and Bob's station are the same.

\begin{figure}[tbp]
\begin{center}
\includegraphics[width=80mm]{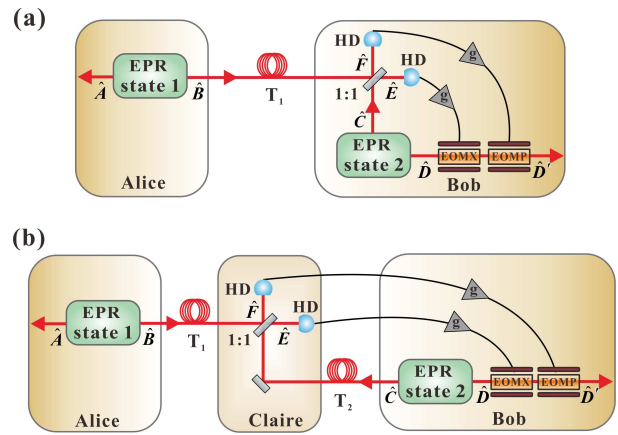}
\end{center}
\caption{The schematic of two swapping schemes of EPR steering. Alice and
Bob own EPR entangled state $\left( \hat{A},\hat{B}\right) $ and $\left( 
\hat{C},\hat{D}\right) $, respectively. (a) Single-channel scheme. A joint
measurement is performed at Bob's station and the measurement results are
fed forward to mode $\hat{D}$. (b) Dual-channel scheme. A joint measurement
is performed at Claire's station. HD: homodyne detector; EOMX and EOMP:
amplitude and phase electro-optical modulators; $T_{1}$, $T_{2}$:
transmission efficiencies of the quantum channels.}
\end{figure}

The amplitude and phase quadratures of an optical mode $\hat{o}$ are defined
as $\hat{x}_{o}$ $=\hat{o}+\hat{o}^{\dagger }$ and $\hat{p}_{o}$ $=(\hat{o}-%
\hat{o}^{\dagger })/i$, respectively. Under this notation, the variances of
the amplitude and phase quadratures for a vacuum state are $V(\hat{x}_{0})=V(%
\hat{p}_{0})=1$, where the subscript 0 represents the vacuum state. A
Gaussian EPR entangled state with a variance $V=\cosh 2r$, where $r\in
\lbrack 0,\infty )$ is the squeezing parameter, can be described by its
covariance matrix 
\begin{equation}
\sigma _{AB}=\left( 
\begin{array}{cc}
V\mathbf{I} & \sqrt{V^{2}-1}\mathbf{Z} \\ 
\sqrt{V^{2}-1}\mathbf{Z} & V\mathbf{I}%
\end{array}%
\right) \text{,}
\end{equation}%
with the matrix element $\sigma _{ij}=\langle \hat{\xi}_{i}\hat{\xi}_{j}+%
\hat{\xi}_{j}\hat{\xi}_{i}\rangle /2-\langle \hat{\xi}_{i}\rangle \langle 
\hat{\xi}_{j}\rangle $, where $\hat{\xi}\equiv (\hat{x}_{A},\hat{p}_{A},\hat{%
x}_{B},\hat{p}_{B})$ is the vector of the field quadratures, and $\mathbf{I}$
and $\mathbf{Z}$ are the Pauli matrices 
\begin{equation}
\mathbf{I=}%
\begin{pmatrix}
1 & 0 \\ 
0 & 1%
\end{pmatrix}%
,\qquad \mathbf{Z}=%
\begin{pmatrix}
1 & 0 \\ 
0 & -1%
\end{pmatrix}%
,
\end{equation}%
respectively.

It has been shown that the steerability of a two-mode Gaussian state can be
easily quantified by \cite{Kogias2015}%
\begin{equation}
\mathcal{G}^{A\rightarrow B}(\sigma _{AB})=\max \left\{ 0,\frac{1}{2}\ln 
\frac{\det \sigma _{A}}{\det \sigma _{AB}}\right\} ,  \label{A}
\end{equation}%
where $\sigma _{A}$ and $\sigma _{AB}$ are the covariance matrices
corresponding to Alice's state and the two-mode Gaussian state,
respectively. $\mathcal{G}^{A\rightarrow B}(\sigma _{AB})>0$ represents that
Alice has the ability to steer Bob's state. Similarly, we have 
\begin{equation}
\mathcal{G}^{B\rightarrow A}(\sigma _{AB})=\max \left\{ 0,\frac{1}{2}\ln 
\frac{\det \sigma _{B}}{\det \sigma _{AB}}\right\} ,  \label{B}
\end{equation}%
which represents Bob's ability to steer Alice's state, where $\sigma _{B}$
is the covariance matrix of Bob's state.

In the ideal case, i.e., with unit transmission efficiency and detection
efficiency, the output modes from the 1:1 beam splitter are $\hat{E}=(\hat{B}%
-\hat{C})/\sqrt{2}$ and $\hat{F}=(\hat{B}+\hat{C})/\sqrt{2}$, respectively.
The amplitude and phase quadratures of the optical modes $\hat{E}$ and $\hat{%
F}$ are measured by two homodyne detectors, respectively. The measurement
results for the optical modes $\hat{E}$ and $\hat{F}$ are represented by $%
\hat{\imath}_{E}=(\hat{x}_{B}-\hat{x}_{C})/\sqrt{2}$ and $\hat{\imath}_{F}=(%
\hat{p}_{B}+\hat{p}_{C})/\sqrt{2}$, respectively. The measured results are
fed forward to optical mode $\hat{D}$ through the classical channels,
respectively. The output beam is%
\begin{equation}
\hat{D}^{\prime }=\hat{D}\ +\sqrt{2}g\ \hat{\imath}_{E}+i\sqrt{2}g\ \hat{%
\imath}_{F},
\end{equation}%
where $g$ describes the amplitude and phase gain factor in the classical
channels, and here we have assumed that the gains in the two classical
channels are equal.

The covariance matrix of the output states $\hat{A}$ and $\hat{D}^{\prime }$
is given by 
\begin{equation}
\sigma _{\text{out}}=\left( 
\begin{array}{cc}
A\mathbf{I} & C\mathbf{Z} \\ 
C\mathbf{Z} & B\mathbf{I}%
\end{array}%
\right) ,  \label{out}
\end{equation}%
where $A=V$, $B=(1+2g^{2})V-2g\sqrt{V^{2}-1}$, and $C=g\sqrt{V^{2}-1}$.
Substituting the matrix elements in Eq. (\ref{out}) into Eqs. (\ref{A}) and (%
\ref{B}), the EPR steering $\mathcal{G}^{A\rightarrow D^{\prime }}$ and $%
\mathcal{G}^{D^{\prime }\rightarrow A}$ can be obtained.

The gain factor in the classical channel is an important parameter in
entanglement swapping. Steerabilities between modes $\hat{A}$ and $\hat{D}%
^{\prime }$ also depend on the gains in the classical channels. The optimal
gains can be obtained by maximizing the steerabilities $\mathcal{G}%
^{A\rightarrow D^{\prime }}$ and $\mathcal{G}^{D^{\prime }\rightarrow A}$,
which are given by 
\begin{eqnarray}
g_{\text{opt}}^{A\rightarrow D^{\prime }} &=&\frac{V\sqrt{V^{2}-1}}{V^{2}+1},
\\
g_{\text{opt}}^{D^{\prime }\rightarrow A} &=&\frac{V}{\sqrt{V^{2}-1}}.
\end{eqnarray}%
respectively. From the expression of $g_{\text{opt}}^{A\rightarrow D^{\prime
}}$ and $g_{\text{opt}}^{D^{\prime }\rightarrow A}$, we see that $g_{\text{%
opt}}^{A\rightarrow D^{\prime }}$ and $g_{\text{opt}}^{D^{\prime
}\rightarrow A}$ are smaller and larger than 1 for any $V$, respectively.
They gradually approach 1 from opposite directions as $V$ increases. In the
limit of $V\rightarrow \infty $, which corresponds to perfect EPR
entanglement, $g_{\text{opt}}^{A\rightarrow D^{\prime }}$ and $g_{\text{opt}%
}^{D^{\prime }\rightarrow A}$ are both equal to 1. The difference between
Eqs. (7) and (8) comes from the asymmetric property of the output state,
i.e., the submatrixes $A\mathbf{I}$ and $B\mathbf{I}$ are different in Eq.
(6) even for the ideal case. In the ideal case, i.e., with unit transmission
efficiency, detection efficiency, and unit gain in the classical channel, we
have $B=3V-2\sqrt{V^{2}-1}$ in Eq. (6). It is obvious that $B$\ is equal to $%
A$\ only when $V$\ is infinite, which corresponds to an infinite squeezing
level. However, an infinite squeezing level is impossible because infinite
pumping power is required. In realistic cases, modes $\hat{B}\ $or/and $\hat{%
C}$\ suffer from transmission losses, so $B$\ and $A$\ are not equal even if
perfect detection efficiency and unit gain are taken.

Loss and noise in quantum channels can lead to decoherence of the quantum
state. Especially, excess noise in the quantum channel, which is the noise
above the vacuum noise, can lead to the disappearance of squeezing and the
sudden death of entanglement \cite{Deng2016,Su20131}. Here, we consider the
proposed swapping schemes for Gaussian EPR steering in lossy and noisy
channels. The losses in the quantum channels are modeled by beam splitters
with transmission efficiencies $T_{1}$ and $T_{2}$, respectively. The excess
noise is modeled by environmental thermal states with noise $W_{1}$ and $%
W_{2}$, respectively. $W_{i}=0$ ($i=1,2$) means that there is no excess
noise and only loss exists in the quantum channel. An optical mode $\hat{o}$
turns into $\sqrt{T}\hat{o}+\sqrt{1-T}(\hat{\nu}+\hat{w})$ after it is
transmitted through a lossy and noisy channel, where $\hat{\nu}$ and\ $\hat{w%
}$ represent the vacuum state and thermal state, respectively. The detection
efficiency $\eta $ of the homodyne detector for modes $\hat{E}$ and $\hat{F}$
is also taken into account, which is modeled by a beam splitter with a
transmission efficiency $\eta $. Taking all imperfections into account, the
covariance matrix elements in Eq. (\ref{out}) are $A=V$, $B=V-2g\sqrt{\eta
T_{2}}\sqrt{V^{2}-1}+g^{2}\left\{ 2+\eta \left[ \left( T_{1}+T_{2}\right)
\left( V-1\right) -T_{1}W_{1}-T_{2}W_{2}+W_{1}+W_{2}\right] \right\} ,$ and $%
C=g\sqrt{\eta T_{1}}\sqrt{V^{2}-1}$, respectively. The corresponding optimal
gain $g_{\text{opt}}^{A\rightarrow D^{\prime }}$ in the classical channels
can be expressed analytically as

\begin{widetext}
\begin{equation}
g_{\text{opt}}^{A\rightarrow D^{\prime }}=\frac{V\sqrt{\eta T_{2}}\sqrt{V^{2}-1}}{%
2V+\eta \left\{ V\left[ \left( T_{1}+T_{2}\right) \left( V-1\right)
-T_{1}W_{1}-T_{2}W_{2}+W_{1}+W_{2}\right] -\left( V^{2}-1\right)
T_{1}\right\}}.
\end{equation}%
\end{widetext}

\begin{figure}[tbph]
\begin{center}
\includegraphics[width=80mm]{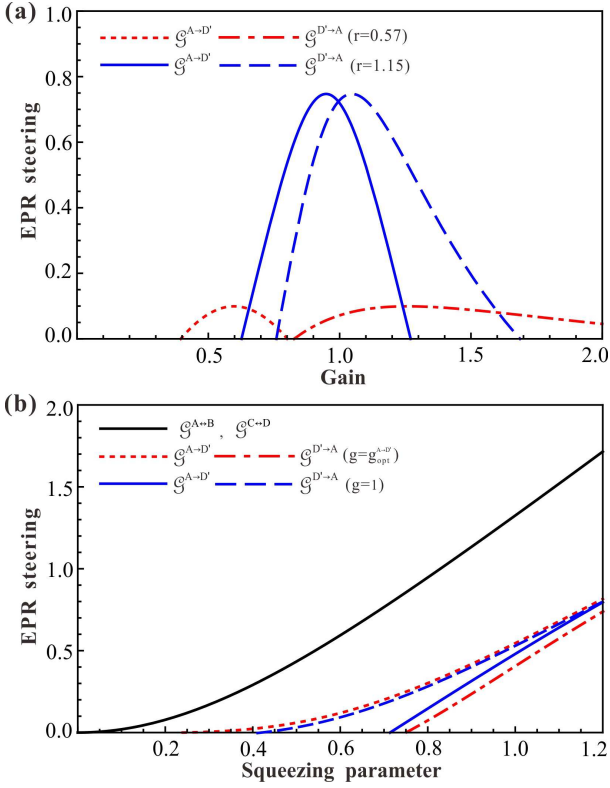}
\end{center}
\caption{(a) Dependence of EPR steerability on gain factors in classical
channels with two different squeezing parameters. (b) Dependence of
steerability on squeezing parameter $r$. The red dotted and dashed-dotted
curves represent the steering between modes $\hat{A}$\ and $\hat{D}^{\prime
} $\textit{\ }when the optimal gain factor $g_{\text{opt}}^{A\rightarrow
D^{\prime }}$ is chosen. The blue\textit{\ }solid and dashed curves
correspond to the steering between $\hat{A}$\ and $\hat{D}^{\prime }$ where
the unit gain factor is chosen. The black curve shows the steerability of
the original EPR resource for comparison.}
\end{figure}

\begin{figure}[tbph]
\begin{center}
\includegraphics[width=80mm]{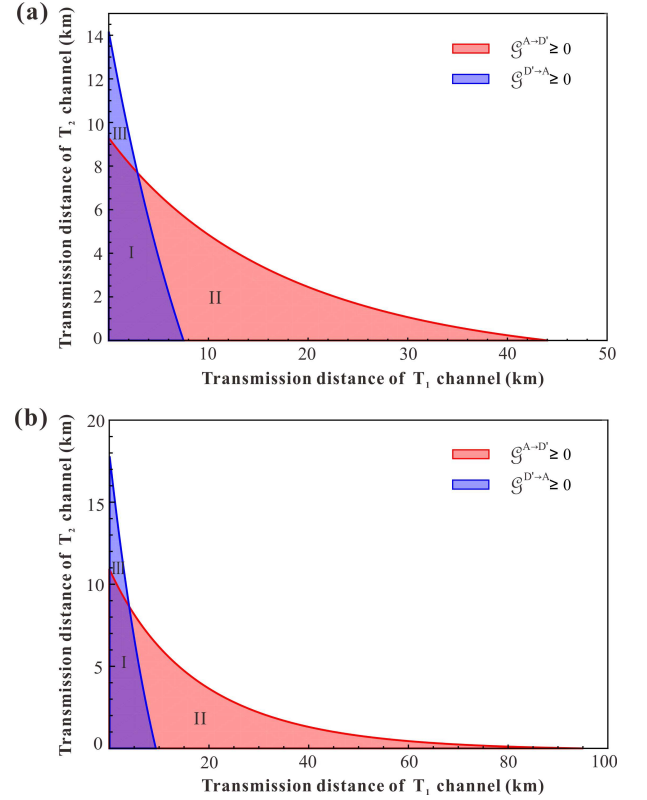}
\end{center}
\caption{The transmission distance regions for output EPR steering in lossy
channels when the optimal gain factor $g_{\text{opt}}^{A\rightarrow
D^{\prime }}$ is chosen. (a) and (b) correspond to a detection efficiency of
95\% and 99.5\%, respectively. Regions I, II, and III correspond to two-way
EPR steering, and one-way EPR steering for $\mathcal{G}^{A\rightarrow
D^{\prime }}$ and $\mathcal{G}^{D^{\prime }\rightarrow A}$, respectively.\ }
\end{figure}

\begin{figure}[tbph]
\begin{center}
\includegraphics[width=80mm]{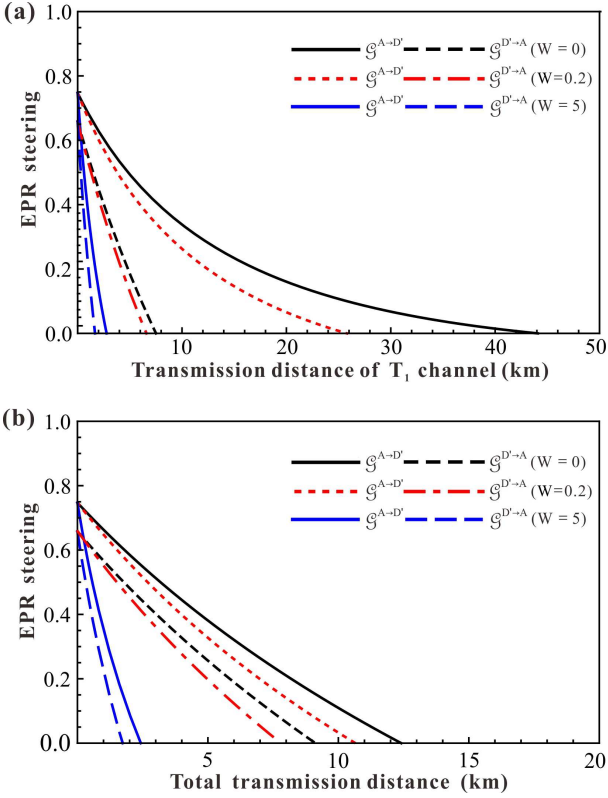}
\end{center}
\caption{Dependence of EPR steering on the transmission distance with
different excess noise for (a) the single-channel scheme and (b) the
symmetric dual-channel scheme. The excess noise is taken as W=0, 0.2, and 5
for comparison. }
\end{figure}

\section{Results and discussions}

Figure 2(a) shows the dependence of output EPR steering on gain factors in
the classical channels with two different squeezing parameters. The
transmission efficiencies $T_{1}$ and $T_{2}$ are chosen as 1, which means
that the swapping process is implemented locally, and the detection
efficiency $\eta $ is chosen as 95\%. Different from entanglement swapping,
the optimal gain factors for $\mathcal{G}^{A\rightarrow D^{\prime }}$ and $%
\mathcal{G}^{D^{\prime }\rightarrow A}$ are different, which is due to the
asymmetry of modes $\hat{A}$ and $\hat{D}^{\prime }$. As $r$ increases, the
optimal gain factors for $\mathcal{G}^{A\rightarrow D^{\prime }}$ and $%
\mathcal{G}^{D^{\prime }\rightarrow A}$ tend to 1, and the maximum
steerabilities also substantially increase. Thus $g=1$ corresponds to the
ideal swapping operation in the limit of infinite squeezing.

Figure 2(b) shows the dependence of output EPR steering on the squeezing
parameter $r$ with optimal gain $g_{\text{opt}}^{A\rightarrow D^{\prime }}$
and unit gain, respectively, where the transmission efficiencies $T_{1}$ and 
$T_{2}$ are chosen as 1 and the detection efficiency $\eta $ is chosen as
95\%. The black curve shows the steerability of the original EPR resource
for comparison. The steerability of modes $\hat{A}$ and $\hat{D}^{\prime }$
is lower than that of the original EPR entangled states $\left( \hat{A},\hat{%
B}\right) $ and $\left( \hat{C},\hat{D}\right) $. When the unit gain is
chosen, EPR steering $\mathcal{G}^{A\rightarrow D^{\prime }}$ and $\mathcal{G%
}^{D^{\prime }\rightarrow A}$ exist only when the squeezing parameter $r$ is
larger than $0.72$ and $0.42$, respectively (blue solid and dashed curves).
One-way steering $\mathcal{G}^{D^{\prime }\rightarrow A}$ is observed in the
range $0.42<r<0.72$. When the optimal gain $g_{\text{opt}}^{A\rightarrow
D^{\prime }}$ is chosen, EPR steering $\mathcal{G}^{A\rightarrow D^{\prime
}} $ and $\mathcal{G}^{D^{\prime }\rightarrow A}$ can be obtained when the
squeezing parameter is larger than $0.24$ and $0.75$, respectively (red
dotted and dashed-dotted curves). One-way steering $\mathcal{G}%
^{A\rightarrow D^{\prime }}$ is observed in the range $0.24<r<0.75$. As the
squeezing parameter $r$ increases, these four curves tend to overlap each
other. Please note that although the optical mode is not transmitted over a
lossy channel in this case, one-way EPR steering is also presented. This is
because the symmetry of the output state is broken after the swapping
process, just as the previous observed one-way EPR steering in a lossy
channel \cite{OneWayNatPhot}.

Comparing the EPR steering $\mathcal{G}^{A\rightarrow D^{\prime }}$ with
unit gain and optimal gain $g_{\text{opt}}^{A\rightarrow D^{\prime }}$, the
required squeezing parameter for $\mathcal{G}^{A\rightarrow D^{\prime }}$ is
reduced from 0.72 to 0.24 with optimal gain $g_{\text{opt}}^{A\rightarrow
D^{\prime }}$. Comparing the EPR steering $\mathcal{G}^{D^{\prime
}\rightarrow A}$ with unit gain and optimal gain $g_{\text{opt}%
}^{A\rightarrow D^{\prime }}$, the required squeezing parameter for $%
\mathcal{G}^{D^{\prime }\rightarrow A}$ is increased from 0.42 to 0.75 by
choosing the optimal gain $g_{\text{opt}}^{A\rightarrow D^{\prime }}$. This
is because the optimal gain $g_{\text{opt}}^{A\rightarrow D^{\prime }}$ is
the maximization of steerability $\mathcal{G}^{A\rightarrow D^{\prime }}$.
If we choose the optimal gain $g_{\text{opt}}^{D^{\prime }\rightarrow A}$,
the required squeezing parameter for $\mathcal{G}^{D^{\prime }\rightarrow A}$
will be reduced while that for $\mathcal{G}^{A\rightarrow D^{\prime }}$ will
be increased. The physical reason for this phenomenon comes from the
asymmetric property of EPR steering. Here, we choose the optimal gain factor 
$g_{\text{opt}}^{A\rightarrow D^{\prime }}$ as an example to present the
results.

Figure 3 shows the transmission distance regions for EPR steering in lossy
but noiseless quantum channels, where the gain factor in the classical
channel is taken as $g_{\text{opt}}^{A\rightarrow D^{\prime }}$. Here, we
consider the transmission loss $\alpha =0.2$ dB/km in the telecommunication
fiber and the squeezing parameter $r=1.15$ (corresponding to 10 dB
squeezing). The detection efficiency is chosen as 95\% and 99.5\% in Figs.
3(a) and 3(b), respectively. In the single-channel scheme [see the
transmission distance of the $T_{1}$ channel in Fig. 3(a)], the maximum
transmission distances for EPR steering $\mathcal{G}^{A\rightarrow D^{\prime
}}$ and $\mathcal{G}^{D^{\prime }\rightarrow A}$ are 45 and 7.6 km,
respectively. When the detection efficiency is improved to 99.5\%, the
transmission distances for EPR steering $\mathcal{G}^{A\rightarrow D^{\prime
}}$ and $\mathcal{G}^{D^{\prime }\rightarrow A}$ in the single-channel
scheme can be increased to 95 and 9.5 km, respectively [Fig. 3(b)]. The
longer transmission distance for $\mathcal{G}^{A\rightarrow D^{\prime }}$ is
obtained because the gain factor is taken as $g_{\text{opt}}^{A\rightarrow
D^{\prime }}$. We also see that the transmission distance of the
single-channel scheme is much longer than that of the symmetric dual-channel
scheme.

With the increase of transmission distance in the quantum channels, two-way
EPR steering (region I) can be turned to either one-way EPR steering $%
\mathcal{G}^{A\rightarrow D^{\prime }}$ (region II) or $\mathcal{G}%
^{D^{\prime }\rightarrow A}$ (region III), respectively. The direction of
one-way EPR steering can be changed at the crossover point of the two
boundary curves for $\mathcal{G}^{A\rightarrow D^{\prime }}\geq 0$ and $%
\mathcal{G}^{D^{\prime }\rightarrow A}\geq 0$ in the dual-channel scheme. As
shown in Fig. 3(a), one-way EPR steering $\mathcal{G}^{D^{\prime
}\rightarrow A}$ can be observed when the transmission distance of mode $%
\hat{B}$ is shorter than 2.9 km (region III), while the one-way EPR steering 
$\mathcal{G}^{A\rightarrow D^{\prime }}$ can be obtained when the
transmission distance of mode $\hat{B}$ is longer than 2.9 km (region II).

Here, we explain the physical reason for the change of the EPR steering
direction. The EPR steering regions I, II, and III in Fig. 3 correspond to
the results of $\mathcal{G}^{A\rightarrow D^{\prime }}>0$ and $\mathcal{G}%
^{D^{\prime }\rightarrow A}>0$, $\mathcal{G}^{A\rightarrow D^{\prime }}>0>%
\mathcal{G}^{D^{\prime }\rightarrow A}$, and $\mathcal{G}^{D^{\prime
}\rightarrow A}>0>\mathcal{G}^{A\rightarrow D^{\prime }}$, respectively.
From the expression of steerabilities in Eqs. (\ref{A}) and (\ref{B}), it
can be clearly seen that the conditions corresponding to the EPR steering
regions I, II, and III in Fig. 3 are $\det \sigma _{A}$ and $\det \sigma
_{D^{\prime }}>\det \sigma _{AD^{\prime }}$, $\det \sigma _{A}>\det \sigma
_{AD^{\prime }}>\det \sigma _{D^{\prime }}$, and $\det \sigma _{D^{\prime
}}>\det \sigma _{AD^{\prime }}>\det \sigma _{A}$, respectively. Two-way EPR
steering can be transformed to one-way EPR steering $A\rightarrow D^{\prime
} $ ($D^{\prime }\rightarrow A$) if the asymmetry of the state exceeds the
boundary $\det \sigma _{AD^{\prime }}=\det \sigma _{D^{\prime }}$ ($\det
\sigma _{AD^{\prime }}=\det \sigma _{A}$) between regions I and II (I and
III). However, it must be pointed out that the asymmetric property of the
two-mode quantum state is only a necessary condition for one-way EPR
steering. In other words, a two-mode quantum state exhibiting one-way EPR
steering must be an asymmetric state, while a two-mode quantum state
exhibiting two-way EPR steering may also be an asymmetric state.

The dependence of EPR steering on the transmission distance in noisy
channels is shown in Fig. 4. Figures 4(a) and 4(b) show the single-channel
scheme and symmetric dual-channel scheme, respectively. The squeezing
parameter $r=1.15$ is chosen, and the excess noise $W$ is taken as $0$, $0.2$%
, and $5$ (in units of shot-noise level), respectively. The optimal gain
factor $g_{\text{opt}}^{A\rightarrow D^{\prime }}$ in the classical channel
is chosen, thus the steerability $\mathcal{G}^{A\rightarrow D^{\prime }}$ is
always larger than $\mathcal{G}^{D^{\prime }\rightarrow A}$ at each noise
level. For simplification, the distances and noise in the two quantum
channels are chosen to be equal in the dual-channel scheme. It is obvious
that the transmission distance in the single-channel scheme is much longer
than that of the symmetric dual-channel scheme at the same excess noise
level. The transmission distances decrease dramatically as excess noise
increases in both schemes. The sudden death of EPR steering can occur when
there is larger excess noise in the quantum channel.

\section{Conclusion}

Comparing with the Gaussian entanglement swapping scheme \cite%
{Ralph,Tan1999,Loock1999,Zhang2002}, the same procedure is used in the
presented swapping schemes of Gaussian EPR steering. There are two
differences between the swapping of Gaussian EPR steering and Gaussian
entanglement. First, in the swapping of Gaussian EPR steering, the obtained
steerabilities of the two remote modes are asymmetric, while the obtained
entanglement of the two remote modes are the same in Gaussian entanglement
swapping. Second, the dependence on the squeezing parameter is different
when the optimal gain is chosen. When the optimal gain in the classical
channel is chosen, higher squeezing is required to complete the swapping of
Gaussian EPR steering, while Gaussian entanglement swapping can be completed
with nonzero squeezing \cite{Zhang2002}.

In conclusion, two swapping schemes of Gaussian EPR steering, a
single-channel scheme and dual-channel scheme, are presented. EPR steering
is observed between two independent quantum modes without a direct
interaction by using the technique of entanglement swapping. The
transmission distances of the single-channel scheme and dual-channel scheme
are compared, and the maximum transmission distance can be obtained by using
the single-channel scheme. The transmission distances are limited by the
squeezing of the Gaussian EPR state and the detection efficiency of the
homodyne detector in the joint measurement. If an EPR state with higher
squeezing and a homodyne detector with higher detection efficiency are used,
a longer transmission distance can be obtained.

One-way EPR steering is presented after the swapping, which is an inherent
property of EPR steering. The change in the EPR steering direction is
observed in the dual-channel scheme, which is related to the asymmetric
property of the output state. In noisy quantum channels, the transmission
distance decreases dramatically with an increase of excess noise in quantum
channels. The presented schemes can be applied in quantum communication
networks with EPR steering.

\section*{ACKNOWLEDGMENT}

This research was supported by the NSFC (Grant No. 61475092, No. 11522433,
and No. 61601270), the program of Youth Sanjin Scholar, the Applied Basic
Research Program of Shanxi province (Grant No. 201601D202006), and National
Basic Research Program of China (Grant No. 2016YFA0301402).

\end{document}